\documentclass[prl,aps,amssymb,twocolumn,showpacs,floatfix]{revtex4}
\usepackage{graphicx}
\usepackage{epsfig}
\usepackage{url}
\usepackage{dcolumn}
\usepackage{bm}

\begin{document}

\title{Dynamical scaling of $YBa_2Cu_3O_{7-\delta}$ thin film conductivity in zero field}

\author{Hua Xu$^{1,}$\footnote[0$^{a)}$]{$^{a)}$email: umdxuhua@gmail.com}}
\author{Su Li$^1$}
\author{C.~J. Lobb$^{1,2}$}
\author{Steven M. Anlage$^{1}$}
\affiliation{$^{1}$Center for Nanophysics and Advanced Materials,
$^{2}$Joint Quantum Institute,\\ Department of Physics, University
of Maryland, College Park, MD 20742-4111}

\begin{abstract}
We study dynamic fluctuation effects of $YBa_2Cu_3O_{7-\delta}$ thin
films in zero field around $T_c$ by doing frequency-dependent
microwave conductivity measurements at different powers. The length
scales probed in the experiments are varied systematically allowing
us to analyze data which are not affected by the finite thickness of
the films, and to observe single-parameter scaling.  DC
current-voltage characteristics have also been measured to
independently probe fluctuations in the same samples. The
combination of DC and microwave measurements allows us to precisely
determine critical parameters. Our results give a dynamical scaling
exponent $z=1.55\pm0.15$, which is consistent with model E-dynamics.
\end{abstract}

 \maketitle

\section{Introduction}
Since the discovery of the high-temperature superconductors, the
normal-superconducting phase transition has attracted great
interest, partially due to the larger critical region which comes
from the high critical temperature and short coherence length of
these materials.\cite{chris,MSalamonAll} There has been a great deal
of work investigating the phase transition in both zero and non-zero
magnetic field.

In zero field, measurements of penetration
depth,\cite{SKamal,SMAnlage} magnetic
susceptibility,\cite{MBSalamon,APomar,RLiang} specific
heat\cite{MBSalamon,Overend} and thermal expansivity\cite{VPasler}
largely agree that the static critical exponent $\nu\simeq0.67$
(governing the divergence of the correlation length
$\xi\sim|T/T_c-1|^{-\nu}$), and indicate that the phase transition
in zero field belongs to the 3D-XY universality class. The dynamics
of the transition, measured through transport properties, remains
uncertain.

In principle conductivity measurements, which depend on the nature
of the dynamics of the order parameter near $T_c$, can determine
both the static critical exponent $\nu$ and dynamical critical
exponent $z$ governing the divergence of the fluctuation lifetime
($\tau\sim\xi^z$). The exponents $\nu$ and $z$ are expected to be
universal, but values extracted from conductivity measurements are
not consistent. For example, DC conductivity measurements yield a
wide range of values for critical exponents: $z = 1.25$ to
$8.3$.\cite{Yeh,Moloni,Roberts,Voss-deHaan}

AC measurements can determine both the real and imaginary parts of
the fluctuation conductivity, providing a stringent test of critical
dynamics.\cite{Jamesbooth,AlanDorsey,Peligrad,ffh} Measurements over
a broad frequency range allow us to probe the dynamical behavior of
the system and directly measure the fluctuation
lifetime.\cite{Jamesbooth} However these experiments are difficult
and seldom done, and the available results are inconsistent, with
values of $z$ ranging from 2 to 5.6. Booth \emph{et al.}
investigated the frequency-dependent microwave conductivity of
$YBa_2Cu_3O_{7-\delta}$(YBCO) films above $T_c$ and obtained $z=2.3$
to $3$.\cite{Jamesbooth} Nakielski \emph{et al.} measured the
conductivity of YBCO at low frequency($<$2 GHz) and obtained
$z\approx5.6$.\cite{Nakielski} Osborn \emph{et al.} did a similar
experiment on $Bi_2Sr_2CaCu_2O_{8+\delta}$ and obtained
$z\approx2$.\cite{KDOsborn} For $La_{2-x}Sr_xCuO_4$, Kitano \emph{et
al.} found $\nu\approx0.67,z\approx2$.\cite{Kiano} There is clearly
a lack of consensus on the dynamic fluctuations of the
superconducting transition. Theory also lacks consensus, with strong
arguments made for both $z=1.5$\cite{Weber,Lidmar,Nogueira} and
$z=2$.\cite{Aji} Further work to extract a value for $z$, and to
establish its validity, is necessary.

When measuring microwave conductivity, one needs to apply a non-zero
current. How the applied microwave current density affects the
measured conductivity has not been systematically addressed.
Recently Sullivan \emph{et al.} argued that a finite-size effect at
low current density was the reason for previous inconsistent results
in DC measurements.\cite{Mattfinitesize} The question of whether the
finite-size effect influences the AC measurement and the extracted
critical exponents, as in DC measurements, inspires us to study the
power dependence of the microwave fluctuation conductivity. We find
that several length scales play a role in AC conductivity
measurements, and only after their effects are properly accounted
for can the underlying critical dynamics be understood.

Our samples are YBCO films($d=$100 nm to 300 nm thickness) deposited
via pulsed laser deposition. AC susceptibility showed $T_c$ of the
films around 90 K with transition widths about 0.2 K. The
resistivity of the films is about $120\mu\Omega$-cm at 2 K above
$T_c$. The fluctuation conductivity was measured by a Corbino
reflection
technique.\cite{Jamesbooth,BoothRSI,APLshielding,xuhthesis}

\begin{figure}
\epsfxsize=3in \epsfysize4in \epsffile{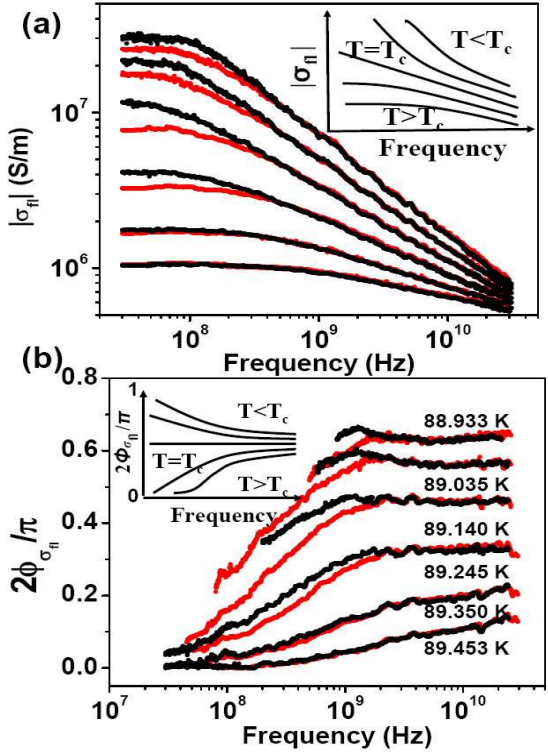} \caption{(Color
online) (a) Magnitude $|\sigma_{fl}|$ and (b) phase
$\phi_{\sigma_{fl}}$ vs. frequency at various temperatures for a
typical YBCO film(xuh139). The blue lines were measured with -22dBm
power while the red lines were measured with -2dBm at the same
temperature. The insets show sketches of Fisher-Fisher-Huse AC
scaling.\cite{ffh}}\label{fig:powercompare}
\end{figure}

Fig.\ \ref{fig:powercompare} shows the measured complex fluctuation
conductivity vs. frequency for various temperatures at two different
microwave powers. The insets in Fig.\ \ref{fig:powercompare} sketch
the expected Fisher-Fisher-Huse(FFH) AC scaling behavior of the
fluctuation conductivity near $T_c$.\cite{ffh}

Fig.\ \ref{fig:powercompare} shows that the measured fluctuation
conductivity has significant and systematic deviations from FFH
scaling. At high frequency, both magnitude and phase of the
fluctuation conductivity look similar to the ideal sketch. However,
as frequency decreases the measured magnitude of the fluctuation
conductivity below $T_c$ saturates, instead of bending up. All of
the phase isotherms below $T_c$ tend toward zero, indicating ohmic
response, instead of approaching $\pi/2$ at low frequency. These
deviations are qualitatively similar to the low current-density
deviations of E vs. J in DC measurements.\cite{Mattfinitesize}

Fig.\ \ref{fig:powercompare} also shows that the applied microwave
power affects the measured fluctuation conductivity, particularly at
low frequencies. As frequency decreases, the higher applied
microwave power decreases the magnitude of the fluctuation
conductivity and depresses the phase. These phenomena can be
explained using scaling theory. Since the critical point is located
in the limit of zero magnetic field $H$, current density $J$ and
frequency $\omega$, increased applied current should drive the
system further away from the transition and thus into the ohmic
regime. The FFH dynamic scaling function can be written in the
following form with assumed dimensionality $D=3$\cite{ffh}:
\begin{eqnarray}
\frac{E}{J} = \xi^{1-z}\chi_{\pm}(J\xi^2,\omega\xi^z,H\xi^2,...).
\label{Eq:FFHGeneral}
\end{eqnarray}
where $E$ is the electric field.

In our measurement, the magnetic field term $H\xi^2$ can be ignored.
The two remaining terms are $J\xi^2$ and $\omega\xi^z$.
Qualitatively, at low frequency, $\omega\xi^z$ is small compared to
$J\xi^2$ so that the applied power has more effect on the
fluctuation conductivity.

\begin{figure}
\begin{center}
\epsfxsize=3in \epsfysize=2in \epsffile{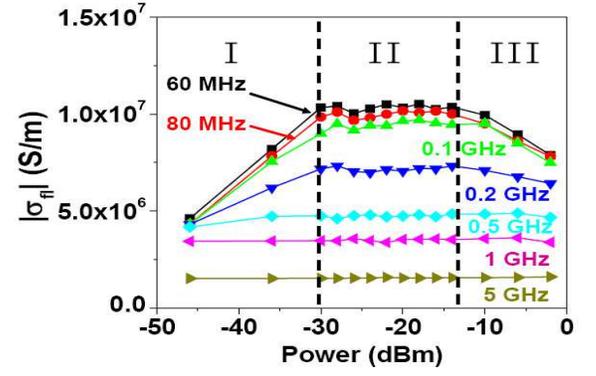}
\end{center}
\caption{(Color online) $|\sigma_{fl}|$ vs. incident microwave power
at different frequencies. ( T=89.140 K, sample xuh139 below
$T_c$)}\label{fig:powerdependence}
\end{figure}

To illustrate the effect of different powers, $|\sigma_{fl}|$ vs.
microwave power at different frequencies is plotted in Fig.\
\ref{fig:powerdependence}. The power dependence of $|\sigma_{fl}|$
varies with frequency. At low frequencies(60MHz, 80MHz and 100MHz),
$|\sigma_{fl}|$ vs. incident power increases first as power
increases(Region I) and then saturates(Region II). At very high
power, the $|\sigma_{fl}|$ decreases again(Region III). At high
frequencies($>0.5$GHz) the fluctuation conductivity is almost power
independent.

The important features in Fig.\ \ref{fig:powerdependence} are that
large applied power affects the fluctuation conductivity, and that
even small power depresses the fluctuation conductivity at low
frequency. While the high-frequency and high-power data in Fig.\
\ref{fig:powerdependence} can be explained by FFH scaling
theory\cite{AlanDorsey,Peligrad,ffh}, the low-power low-frequency
behavior is not explained by these theories alone.

The similarity between this low power and low frequency deviation
and the low current density deviation in DC conductivity measurement
suggests the presence of a "probed length scale" for a finite
frequency. In a simple physical model, fluctuations can be
visualized as closed circular vortex loops of radius r. In an
infinite superconductor with no applied current, vortex loops of
different size occur with different probabilities as thermal
fluctuations. When a current with density J is applied, some vortex
loops(with large r) will blow out to infinite size(dissipation).
Some vortex loops(with small r) shrink and annihilate(no
dissipation). The current density induced length scale $L_J$, which
can be written as $L_J = (\frac{k T}{2 \pi \Phi_0
J})^{\frac{1}{2}}$,\cite{ffh,Mattfinitesize,Woltgens} separates
vortex loops into two categories, depending on their ultimate fate.
The shrinking of a vortex takes time. The shrinking time depends on
the size of the vortex, thus relating the size of a vortex to a time
scale. In AC measurements, small frequency means that large length
scales are probed, so one investigates large size vortex loops, and
\emph{vice versa}. From the order parameter relaxation time scale in
time-dependent Ginzburg Landau theory\cite{Tinkham} one can
construct a frequency-induced length scale $L_{\omega} =
(\frac{ck_BT_c}{\hbar\omega})^{1/z}\xi(0)$, where $c$ is a constant
of order 1 and $\xi(0)/\xi(T)=|T/T_c-1|^{\nu}$.

In AC conductivity measurements, the probed length scale involves
both frequency and current density. Among them, the smaller length
scale dominates the measured fluctuation conductivity. We propose a
plausible expression for the probed length scale for AC measurement
$L_{AC}$, $\frac{1}{L_{AC}} = \frac{1}{L_J}+\frac{1}{L_{\omega}}$.
This formula has the correct limits, and it is also consistent with
the two-term FFH scaling, without the magnetic term $H\xi^2$. Finite
size effects come into play when $L_{AC}$ approaches the thickness
of the film.

\begin{figure}
\begin{center}
\epsfxsize=3.2in \epsfysize=2.2in \epsffile{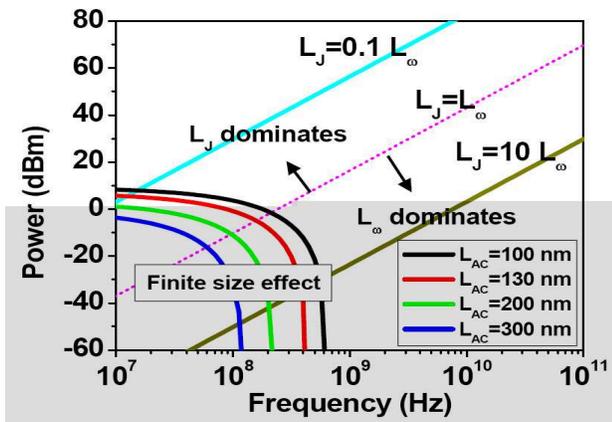}
\end{center}
\caption{(Color online) Summary of length scales and finite size
effects in Corbino AC measurements of fluctuation conductivity of
YBCO films near $T_c$. The dotted line in the figure gives the
boundary $L_J = L_{\omega}$. At low frequency and small current
density, the probed length scale $L_{AC}$ approaches the thickness
of the sample. }\label{fig:AClengthscales}
\end{figure}

Fig.\ \ref{fig:AClengthscales} summarizes the length scales in an AC
measurement in terms of experimental quantities. In this figure, we
use $\xi(0)=5${\AA}, $c=1$ and $z=1.5$. The dotted line in the
figure gives the boundary $L_J = L_{\omega}$. To the right and below
the dotted line, when $L_{\omega}\ll L_J$, the frequency induced
length scale dominates, and one observes mainly frequency dependent
behavior of the fluctuation conductivity. Above the dotted line,
when $L_{\omega}\gg L_J$, current-induced nonlinear effects will
dominate the behavior. This explains the features shown in Fig.\
\ref{fig:powercompare} and Fig.\ \ref{fig:powerdependence}, where
the current density has less effect on the fluctuation conductivity
at high frequency and a larger effect at low frequency.

At low frequency and small current density, $L_{AC}$ may approach
the thickness of the sample (\emph{d}) or some other length scale
that interrupts the fluctuation vortex loops. Hence deviations from
the simple scaling theory are expected when $L_{AC}>d$.

In our AC measurements, we want to keep $L_{AC}<d$. Hence we choose
to stay at low $J$ but high $\omega$. In this region we can find the
true critical behavior without getting into finite-size effect or
crossover difficulties. Our previous analysis strayed out of this
region and may account for the larger values of $z$ reported
before\cite{Jamesbooth} and elsewhere in the literature.

With very small applied microwave power, -46dBm, and high frequency
data, we investigated the frequency dependent fluctuation
conductivity around $T_c$. We found that the determination of $T_c$
is crucial for obtaining critical exponents. With high quality data
taken at small temperature intervals (50 mK),\cite{AlanDorsey} first
we improved the conventional data analysis method\cite{Jamesbooth}
to determine $T_c$. We did a quadratic fit for $\log
|\sigma_{fl}(f)|$ vs. $\log(f)$ and a linear fit for
$\phi_{\sigma}(\omega)$ vs. $\log(f)$ and determined $T_c$ through
the sign of the $\log(f)^2$ coefficient of the quadratic fit, and
the slope of the linear fit.\cite{xuhthesis} In addition, we
developed a new method to determine $T_c$. Inspired by the Wickham
and Dorsey scaling function $S_+(y)$\cite{AlanDorsey}, we choose
scaling parameters $\omega_0(T)$ and $\sigma_0(T)$ at each
temperature to collapse $\phi_{\sigma}(T)$ vs. $\omega/\omega_0(T)$
and $|\sigma_{fl}|/\sigma_0(T)$ vs. $\omega/\omega_0(T)$ to smooth
and continuous curves, without \emph{a priori} determination of
$T_c$ or critical exponents.\cite{Kiano,xuhthesis} Then we extracted
the critical temperature and exponents by forcing $\omega_0(T)$ and
$\sigma_0(T)$ to show power-law behavior near $T_c$. By combining
the two methods the critical temperature and exponent for sample
xuh139 were determined, $T_c = 89.25\pm0.03K,z = 1.55\pm
0.12$.\cite{xuhthesis}

The dynamic critical exponent should be sample independent. To check
the results, we not only repeated measurements on the same sample,
but also repeated the experiment on different samples. Films of
different thickness($d=$100 nm to 300 nm) were examined, and z was
found to be independent of the thickness, keeping in mind the
constraints of Fig.\ \ref{fig:AClengthscales}. Experiments on 6
samples have been done giving $z=1.56\pm0.10$.

\begin{figure}
\begin{center}
\epsfxsize=3.2in\epsfysize=3.8in \epsffile{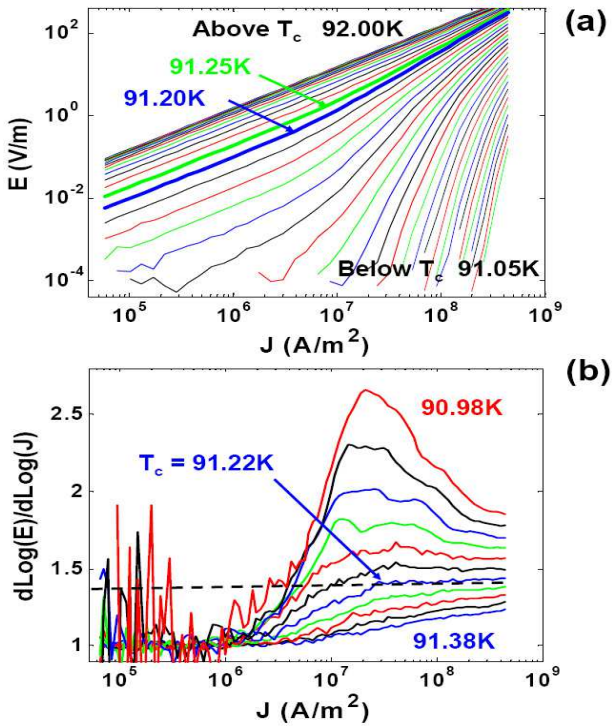}
\end{center}
\caption{(Color online) DC current-voltage characteristics
measurement, performed after the AC experiment on xuh139 in zero
magnetic field. (a) E-J isotherms(50 mK apart), (b) $d\log E/d\log
J$ vs. J derivative plot(40 mK apart).}\label{fig:Figure4}
\end{figure}

We also performed DC current-voltage characteristic measurements on
the same samples.\cite{xuhthesis} Typical results are shown in Fig.\
\ref{fig:Figure4}.(with no background subtraction\cite{Lisuthesis})
According to the negative curvature criterion\cite{doug}, we
determined the critical temperature to be $91.220\pm0.04$ K and the
critical exponent $z=1.75\pm0.2$ from the derivative plot in Fig.\
\ref{fig:Figure4}(b). In Fig.\ \ref{fig:Figure4}, all the isotherms
tend towards ohmic behavior at low current density, brought about by
$L_J>d$ finite-size effects.\cite{Mattfinitesize} When the current
density is smaller than $1\times10^6 A/m^2$, the sample will have
only ohmic response around $T_c$. The -46dBm applied power in AC
measurement corresponds to a maximum current density of
$2.2\times10^5A/m^2$($<1\times10^6A/m^2$). This means that for
-46dBm incident power $L_J>d$, verifying a feature of Fig.\
\ref{fig:AClengthscales}, and suggesting that one-parameter scaling
should work when $L_{\omega}<L_J,d$. Hence it is appropriate to
determine $T_c$ and critical exponents with AC data at  -46dBm
applied power.

The difference of $T_c$ between DC and AC measurements is due to the
different thermometer positions and temperature control techniques
of the two systems. The resistance vs. temperature plots from the AC
and DC experiment have a 2.0 K temperature offset, which is the
difference of the determined $T_c$ from these two methods. Hence the
determined $T_c$ from the two methods are consistent.

Performing DC measurements on the same film after AC measurements
involves more processing steps than a DC measurement alone, and may
result in additional disorder in the sample. Disorder and heating
leads one to systematically choose a lower temperature isotherm as
$T_c$ and then obtain a larger value of
$z$.\cite{xuhthesis,Lisuthesis} We carefully repeated the
measurements on different YBCO films and found that the sample
quality does affect the obtained value of $z$ from the DC
experiment.\cite{xuhthesis} However, for films with high $T_c$,
sharp transition and small resistivity, we obtained a value of
$z=1.56\pm0.08$ from DC measurement, which is consistent with the AC
result $z=1.55\pm0.15$. In addition, DC measurements carried out the
same way on high-quality crystals also gave
$z\approx1.5$.\cite{Lisuthesis}

In this work, we focused on temperatures very close to $T_c$(the
reduced temperature $t<0.004$) and finite-frequency scaling. In this
temperature range, due to the influence of the finite size effect at
low frequencies the static critical exponent $\nu$ cannot be
determined.\cite{Lisuthesis}

To conclude, using two different measurement methods, we studied the
dynamic fluctuation effects of $YBa_2Cu_3O_{7-\delta}$ thin films
around $T_c$. The combination of AC and DC measurements precisely
determined the dynamical scaling exponent $z=1.55\pm0.08$, which
suggests the superconducting to normal phase transition of
high-$T_c$ materials is consistent with model
E-dynamics.\cite{Hohenberg}

The authors thank A. T. Dorsey for insightful discussion. This work
has been supported by NSF grant number DMR-0302596.



\end{document}